\documentclass{article}

\begin{document}

{\large\bf An Evenly-Spaced LSST Cadence for Rapidly Variable Stars}

~\\~\\~\\

Eric D. Feigelson$^{1,2}$,  Federica B. Bianco$^{3,4,5}$ and Rosaria Bonito$^6$

~\\

$^1$ {Department of Astronomy \& Astrophysics, Pennsylvania State University, 525 Davey Laboratory, University Park, PA 16802, USA}

$^2${Center for Astrostatistics, Pennsylvania State University}

$^3$ {Department of Physics and Astronomy, University of Delaware, Newark, DE 19716, USA}

$^4$  {Joseph R. Biden, Jr. School of Public Policy and Administration, University of Delaware, Newark, DE 19716, USA}

$^5$ {Data Science Institute, University of Delaware, Newark, DE 19716, USA}

$^6$ {INAF—Osservatorio Astronomico di Palermo, piazza del Parlamento 1, 90134 Palermo, Italy}

{Accepted for publication in Astrophysical Journal Supplements (July 2023)}

~\\~\\

ABSTRACT

Stars exhibit a bewildering variety of rapidly variable behaviors ranging from explosive magnetic flares to stochastically changing accretion to periodic pulsations or rotation.  The principal Rubin Observatory Legacy Survey of Space and Time (LSST) surveys will have cadences too sparse and irregular to capture many of these phenomena.  We propose here a LSST micro-survey to observe a single Galactic field, rich in unobscured stars, in a continuous sequence of 30 second exposures for one long winter night in a single photometric band.  The result will be a unique dataset of $\sim 1$ million regularly spaced stellar light curves (LCs).  The LCs will constitute a comprehensive collection of late-type stellar flaring, but also other classes like short-period binary systems and cataclysmic variables, young stellar objects and ultra-short period exoplanets.  An unknown variety of anomalous Solar System, Galactic and extragalactic variables and transients may also be present. A powerful array of statistical procedures can be applied to individual LCs from the long-standing fields of time series analysis, signal processing and econometrics.  Dozens of `features' describing the variability can be extracted and the ensemble of light curves can be subject to advanced machine learning clustering procedures.  This will give a unique, authoritative, objective taxonomy of the rapidly variable sky derived from identically cadenced  LCs.  This micro-survey is best performed early in the Rubin Observatory program, and the results can inform the wider community on the best approaches to variable star identification and classification from the sparse, irregular cadences that dominate the planned surveys.

\section{Introduction}

Ground-based astronomical surveys are now in a mature phase with several projects providing hundreds or thousands of photometric observations of many millions of objects.  They were often initiated for specific purposes: OGLE for discovering gravitational microlensing events, ZTF and ASAS-SN for extragalactic transients, WASP and HATnet for transiting exoplanets.  They have also been powerful engines for expanding samples of variable stars particularly periodic variables like Cepheids, contact eclipsing binaries, rotational and pulsational variables (e.g. Soszynski et al. 2008, Mowlav et al. 2018, Chen et al. 2020).  They are also effective at discovering eruptive events such as supernovae and cataclysmic variables.  

But due to their cadence structures, these surveys are less effective at detecting and characterizing variability on rapid timescales of minutes to hours.  For example, a handful of variable stars with periods $<$ 90 minutes are found among $\sim 10$ million variable candidates in the ZTF survey  (Ofek et al. 2020) and a few dozen M dwarf flares are found in four years of ASAS-SN transient alerts (Schmidt et al. 2019).  They miss classes such as double white dwarf binaries with periods down to $\sim 5$ minutes (Kruckow et al. 2021), transiting planets with periods down to $\sim 4$ hours (Smith et al. 2018), magnetic reconnection flares with durations from minutes to hours (Namekata et al. 2017), and stochastic flickering in cataclysmic variable disks with characteristic timescales of around 1000~s (Dobrotka 2021).

While rapid photometric monitoring of single stars has a long history (e.g. Worden 1981), a few recent efforts have been made with rapid observations of stellar populations. The Kepler, K2 and TESS missions have a small fraction of rapid cadence observations (e.g. 1 minute), but these are oriented towards astroseismological studies of individual stars.   Gaia and long-standing ground-based surveys (such as OGLE, ZTF, WASP, and HATnet) have sparse and irregular cadences that are not adapted to study of rapid variability.  A search for visible-band extragalactic transients associated with gravitational wave events, Fast Radio Bursts, gamma-ray bursts, and supernova shock breakouts has been made (Andreoni et al. 2020).   Saha et al. (2019) observed six crowded fields in Baade's Window using the DECam instrument with $5-300$~s exposures over several dozen non-sequential visits, detecting $\sim 5000$ variables out of $\sim 2.5$ millions stars including short-period binaries, $\delta$ Scuti and SX Phe pulsators, and RR Lyrae stars.  However,  the cadence of these observations was $\sim 2$ hours for each night of observing time.  The DECaPS (Dark Energy Camera Plane Survey) time domain survey of two LSST Deep Drilling Fields have been observed with $\sim 1000$ images using a 30 second cadence (Graham et al. 2023).  But these exposures were distributed over several dozen nights with continuous observations lasting only 18 minutes.

The Legacy Survey of Space and Time (LSST) of the Vera C. Rubin Observatory is the most ambitious program in time domain astronomy in the history of ground-based astronomy (Tyson 2002, LSST Science Collaboration 2009, Ivezic et al. 2019).  The great majority of its time will be spend in the Wide Fast Deep (WFD) survey which will give very sparse cadences for each pointing in a given photometric band.  The Deep Drilling Fields (DDFs) will have denser cadences but still with very irregular spacings.  Stellar studies are particularly affected because planned cadences for the Galactic Plane are thin. The cadences are generally oriented towards the discovery and classification of cosmic objects with variations on timescales of days, months and years.  The principal classes that LSST can identify are supernovae, active galactic nuclei, extragalactic transients, and a few types of variables stars. The chosen cadence strategy of the WFD and DDF fields is further compounded by unavoidable diurnal and annual gaps due to solar motion across the sky.  

The resulting sparse and irregular cadences has two important deficiencies. First, the LSST WFD and DDF cadences access only a portion of the the range of variability characteristics of Galactic stars. Periodic binary star orbits range from 11 minutes to centuries.   Omitting neutron stars, stellar pulsation periods range from tens of seconds to months.   Stochastic accretion variations in cataclysmic variables range from milliseconds to decades.   Magnetic flaring occurrences can be nearly continuous or unique. The temporal behaviors have a bewildering variety of periodic, stochastic and explosive behaviors. An {\it ab initio} effort to characterize and classify variable stars from the WFD and DDF observations will be very difficult.  Bonito et al. (2023) demonstrate in detail how these LSST surveys will not adequately characterize the diverse variability behaviors of accreting pre-main sequence stars. 

Second, the methodology for characterizing sparse and irregularly cadenced light curves is limited to a narrow suite of methods developed by astronomers.  The mathematical foundations for statistics applied to these time series are often insecure; for example, a dozen astrostatistical studies have wrestled with  False Alarm Probabilities for Lomb-Scargle periodograms.  It has proved remarkably difficult to ascertain with confidence, for a sparse irregular cadence, the existence of periodic behavior (e.g. Suveges et al. 2015, vanderPlas 2018).

These deficiencies can be partially recuperated by committing a small portion of the Rubin Observatory time to a very different cadence: observing a single field continuously in a single band with a rapid cadence. The resulting dense and evenly spaced cadence of a field with $\sim 1$ million diverse cosmic objects will detect rapid stellar variables, and other transient events, that are poorly covered by the WFD and DDF surveys. It will permit application of a wide range of statistically validated methods from time series analysis tools used in signal processing, econometrics, and biomedicine. These time series methods might `anchor' the  limited suite of methods available for sparse and irregular cadences, allowing improved classification of billions of WFD stars. Powerful new science and methodological results can emerge from a single night of this dense, rapid cadence supplemented by brief multiband photometry of the same field.

These issues are elucidated further in this paper.  Section \ref{rapidsky.sec} summarizes science goals relating to short timescale variability in repeated LSST exposures of the same field.  Sections \ref{method.sec}-\ref{ensemble.sec} discuss the statistical methodology available to dense regularly cadenced, in contrast to sparse irregularly cadenced, observations.  Section \ref{1night.sec} outlines specifics of a proposed single night observation during the LSST Commissioning Phase.  However, the actual pointing will depend on the night allocated for the microsurvey.

\section{The Rapidly Varying Stellar Sky}
\label{rapidsky.sec}

\subsection{Magnetic reconnection flares}

dMe flare stars are, by far, the most common sources of rapid variability in the sky, even at high Galactic latitudes.  The Deep Lens Survey, searching for extragalactic transients with wide-field cameras on 4m-class telescope using rapid cadences, finds that extragalactic transients are lost in a 'foreground fog' of dMe flares with an annual all-sky rate around $\sim 10^8$  (Kulkarni 2006). 

Population characteristics  emerged from analysis of $\sim 10,000$ flare stars in the Sloan Digital Sky Survey Stripe 82 (Kowalski et al. 2009): M4-M6 stars near the Galactic Plane were most commonly seen to flare, although more luminous flares are produced by M0-M1 stars.  Roughly 0.1\% of M4-M6  stars are flaring during a single observation.  Additional results from the Deep, Wide, Fast program using DECam show flare energies over a wide range $log E \sim 31-37$ erg with a frequency distribution $dN/dE \propto E^{-1.4}$ above $logE = 33$~erg (Webb et al. 2021).  The observed occurrence rate is roughly 10~flares~deg$^{-2}$~ day$^{-1}$ but this is strongly dependent on Galactic location and sensitivity.

{\bf Low-mass flaring stars} ~~ High amplitude magnetic reconnection flares on M dwarfs were among the most spectacular explosive events in classical astronomy.  On 12 April 1985, the nearby dM3.5e star AD~Leo with mass $0.5$~M$_\odot$ experienced a flare with $log E \sim 34$~erg,  brightening an order of magnitude in the R band in 4 minutes with flux decaying over the next hour (Hawley et al. 1991).  $Kepler$ and $TESS$ surveys show that active mid-M stars produce flares every day with energies exceeding $log E \sim 32 $~erg and are nearly continuously flaring at energies $> 29$~erg (Hawley et al. 2014, Gunther et al. 2020).  Large samples are needed to unravel the flaring dependence on stellar mass and age, and current samples are deficient in the cooler later than spectral type M5.

These will be the most common variable stars in the rapid cadence LSST single night field.   As the stellar Initial MassFunction peaks around 0.3~M$_\odot$, the majority of stars in the LSST field will be M stars. A study similar to our proposal, but much smaller scale, has recently been conducted with rapidly cadenced $g$ band observations using DECam on the Blanco 4m telescope (Webb et al. 2021).  About 100 flares were detected from a sample of 20,000 relatively bright stars with $Gaia$ distances within 500~pc. Flare energies $log E \sim 31-37$ erg from M0$-$M6 stars, most with timescales $< 8$ minutes.   

The number of expected M dwarf flares during the single night LSST observation can also be roughly estimated from Kepler mission findings.  The 4-year photometric survey with 29-minute cadence designed for transiting planet discovery shows $\simeq 2$\% of M dwarfs each produced $>$100 flares during the 4 year mission (Davenport et al. 2016).  This predicts $\sim 10,000$ M dwarfs will produce a single flare during the 1-night LSST observation.  But this estimate is quite uncertain. First, the Kepler flares typically have brightness ratios $10^{-4} < f_{flare}/f_\star <  10^{-3}$ averaged over flare durations; some will be too faint to be detectable with a ground-based telescope implying fewer than 10,000 detections.  But M dwarfs that flare less frequently than those in the sample of Davenport will increase the predicted rate. 

There are two issues of particular astrophysical interest.  First, main sequence stars undergo a  transition from partially to completely convective interiors around M2-M3 suggesting that the magnetic dynamos may switch from tachoclinal to distributed convective mechanisms.  Evidence from Ca~II line intensities is growing that magnetic activity changes at this transition but no flare studies have yet shown an effect (Mullan et al. 2020). Second, exoplanetary surveys have now established that closely packed systems of rocky planets orbit most M stars with many in the habitable zones (Shields et al. 2016).  But habitability will also depend on the effects of stellar flares and associated coronal mass ejections which may photoevaporate and erode planetary atmospheres on relatively short timescales: the oceans may disappear.  

{\bf Solar-mass flaring stars} ~~ One of the great excitements from the $Kepler$ mission was the discovery of 'superflares' from older solar-type stars (Maehara et al. 2012 and 2015). Together with $TESS$ studies (Tu et al. 21), superflares have been seen from $\sim 500$ stars out of $\sim 100,000$ observed stars.  As with M flares, solar-type flares rise in a few minutes and last for less than an hour.  Energies are mostly in the range $log E \sim 33-36$ erg with occurrence timescales of thousands of years for typical stars.   This is far above levels expected from solar activity, but 5 orders of magnitude below the flare rates found by Webb et al. (2021) for M stars. 

The single night LSST observations are not likely to include more than some tens of thousands of solar-type stars, so the sample of detected superflares from these stars is expected to be very small, fewer than 10  and possibly zero solar-mass superflares.

\subsection{Other known rapid stellar variables}

{\bf Short-period binary systems} ~~ Double-line spectroscopic binaries and photometric eclipsing binaries (EBs) have played a central role in establishing models of stellar physics and evolution for over a century.  The $Kepler$ and OGLE surveys find that short-period EBs with periods $< 0.5$~day have an occurrence rate around 0.2\% (Bienias et al. 2021, Bodiet al. 2021). Systems range from detached to semi-detached and contact binaries with diverse light curve morphologies.   Roughly half have depths 2-8\% and periods 0.25-0.32 day with many other having deeper depths or shorter periods down to 0.18 days (Bienias et al.).   

With 5 mmag expected precision, LSST photometry may identify eclipsing binary (EB) light curves with transit depths as shallow as $\sim 3$\%.  The shapes of EB light curves are so distinctive that only two orbital periods may be needed for reliable classification.  We thus expect to capture a substantial fraction of the hundreds or thousands of short-period EBs in the LSST field.  An unknown, and possibly larger, population of non-eclipsing binaries should be found where periodic variations arise from mutual illumination and tidal distortion of the component stars.  This quasi-sinusoidal variation is seen as curvature between transits in the LCs in eclipsing binaries. 

{\bf White dwarf binary systems} ~~ Several dozen compact and ultra-compact white dwarf binaries, some with periods under 1 hour, have been found.  Some are non-eclipsing and would not show a photometric signature (Brown et al. 2020), but others exhibit high-amplitude eclipses with a variety of morphologies and periods ranging from $\sim 0.1$ to 1 hour (Burdge et al. 2020).  These systems are particularly important as gravitational wave sources potentially detectable with the forthcoming $LISA$ mission. The number expected to lie in a single LSST field is uncertain.  A catalog of ultracompact white dwarf binaries might also be created from the LSST WFD survey with the advantage of a wide-field scope but disadvantage of sparse observing cadence (Breivik et al. 2022, B.6.3.1). 

{\bf Cataclysmic variables}  ~~  The most common stellar compact object is the white dwarf and, in close binary systems, accretion from a normal or giant companion can readily occur for extended periods. Several types of cataclysmic variables (CVs) are excellent testbeds for close binary star evolution. CVs are mostly found at intermediate Galactic latitudes ($15^\circ - 45^\circ$) and are in outburst $5\% - 20\%$ of the time (Drake et al. 2014).  Orbital periods typically range from 1 to 10 hours with a period gap at $2.3-3.4$ hours.

But observed samples are badly incomplete: only  $\sim$10,000 CVs are known from all surveys.  The $CRTS$ and $Gaia$ surveys each report $0.3-1$ CV per day, most with quiescent levels around 20~mag. The space density of CVs smoothed over the Galactic disk height is quite uncertain.  An empirical estimate based on a volume-limited sample gives $5 \times 10^{-6}$ CVs~pc$^{-3}$ (Pala et al. 2020) while  theoretical models predict a much higher density around  $2-10 \times 10^{-5}$~pc$^{-3}$ (Belloni et al. 2018).  

LSST has several times the \'etendue of CRTS and might thus expect to see several bright CVs in a single-night observation.  But the total number of CVs including the fainter stars that dominate the $\sim 1$ million star sample is difficult to estimate given the uncertainties in space density and in the fraction of active to quiescent systems. The number and photometric properties of CVs found in the micro-survey should give improved estimates of these two uncertain quantities.  Most CVs are from systems formed $1-10$~Gyr ago and will therefore populate the upper disk region of the planned target. CVs in their active phase have distinctive rapid flickering that should allow unambiguous identification.

{\bf Young stellar objects}  ~~  While CVs are restricted to a small fraction of close binary systems, all stars pass through the pre-main sequence phase with intense accretion and magnetic activity.  Accretion dominates the  protostellar ($<0.2$ Myr) and classical T Tauri ($< 5$ Myr) phases, while magnetic activity dominates in disk-free phase ($< 30$ Myr).  Photometric variations due to accretion and magnetic activity  are commonly seen at the level of several percent on timescales of hours (Venuti et al. 2021, Bonito et al. 2023).  A wide variety of behaviors are seen:  periodic, quasi-periodic, stochastic, bursting, dipper, eclipsing and secular variations.  Photometric variations with $\sim 10$\% amplitudes on timescales of minutes and hours is sometimes present in accreting T Tauri stars (Gullbring et al. 1994, Smith et al. 1996). 

These stars will be present in the target LSST single night field if it is pointed is near a massive star forming complex like the Carina Nebula.  This complex has $\sim 100,000$ pre-main sequence in the inner 1 square degree, and likely has a halo of previously formed  of many more stars spread over several degrees (Feigelson et al. 2011). The difficulties of characterizing young stellar variability in the sparse WFD survey has been discussed by Bonito et al. (2023). 
 
{\bf Ultra-short period exoplanets} ~~ It is now understood that most stars are orbited by several planets, and often in compact configurations close to the host star.  Ultra-short period (USP) planets have orbital periods shorter than one day.  While most known USPs have Neptune or super-Earth sizes that are not detectable from ground-based photometric surveys, a few are USP hot Jupiters.  These systems are astrophysically important: their atmospheres are inflated and eroding, and their orbits may be rapidly decaying due to tidal dissipation. The currently shortest period is 18 hours where a 1.2~R$_J$ planet is orbiting a 14 mag K star (McCormac et al. 2020). 

These USP Jupiters are rare.  Based on a survey of 1 million stars with the TESS satellite (Melton et al. 2023), we roughly expect $\sim 10$ such systems to be present in this LSST microsurvey. It is possible they will suffer strong contamination by unequal-mass stellar binaries.  

\subsection{Known and unknown unknowns}

In addition to known classes of rapidly variable stars, other transient and variable events may be detected in this proposed LSST microsurvey.  Scientifically uninteresting anthropogenic effects will quite possibly appear, such as satellite glints around dawn and dusk (Schaefer 1987).  If the target field is near the ecliptic plane\footnote{Such as (l,b)=(347, +30) rather than (330, +20) suggested in \S\ref{1night.sec}.}, rotating small ($<$100m) asteroids may show periodic variations on timescales of minutes (Harris et al. 2006). `Orphaned tracklets’ might appear due to rapidly dissipating dust clouds arising from catastrophic collision of small asteroids (LSST Science Collaboration 2009, \S5.6.1). In nonthermal extragalactic active galactic nuclei, optical micro-variations on timescales of minutes to hours have been studied in radio-loud narrow-line Seyfert 1 galaxies (Ojha et al. 2022) and in relativistic blazar jets (Chand 2022).  In stellar astrophysics, there may be magnetic, accretional, collisional, or shock breakout explosive phenomena that we do not know.  

Little is known about explosive transients in the 0.001-0.1 day region. Theoretically, the most promising possibility may be a white dwarf `naked explosion' where the timescale for photons to leak out of the stellar material is $\sim 1000$~sec with peak luminosity $\sim 100$~L$_\odot$ (Kulkarni. 2006).  This phenomenon has never been seen.  A decade ago, Kulkarni (2012) wrote: "Based on the history of [optical transient research] we should not be surprised to find, say a decade from now, that we were not sufficiently imaginative".  The proposed micro-survey will be the deepest investigation of a rarely studied regime of celestial phenomena; remarkable findings may emerge.

\section{Time Series Methodology for Sparse, Irregular Cadences}
\label{method.sec}

The summary above shows that a wide variety of temporal phenomena are expected from stars when accurate photometric measurements are made on timescales of minutes to hours.  From a statistical perspectives, some of these variations are $nonstationary$ with either continuous trends in mean flux or sudden events superposed on a constant flux.  Other phenomena produce $stationary$ variability where the changes are continuous.  These can be periodic variations from orbits or pulsations, or stochastic noise processes (often in combination with deterministic trends). Noise can be simple uncorrelated Gaussian (normal) noise or complex autoregressive stochastic processes.  Statistical methods are thus needed to detect and discriminate stationary and nonstationary, deterministic and stochastic, periodic and aperiodic variations. 

Most time series methodology is designed for engineered systems or human affairs where it is natural to arrange the acquisition of evenly spaced time series.  Astronomers are unusual in their inability to obtain, except with rare instrumentation such as NASA's $Kepler$ and $TESS$ missions, regular cadences.  Interruption in data acquisition are commonly due to diurnal and annual solar motion, lunar cycles, bad observing condition, and competition for time on telescopes that serve many astronomical goals.   

Statistical methods for treating sparse and irregular cadence time series have thus been developed mostly in the astronomical community with little contribution from statisticians or other practitioners.  Examples of methods for these cadences are: moments of the brightness distribution (mean, variances, skewness, kurtosis); the chi-squared and other measures of photometric variability amplitude (e.g. Welch et al. 1993); discrete correlation function to study autocorrelation (Edelson \& Krolik 1988); sigma-clipping to identify outliers (Astropy Collaboration 2018), and various interpolation procedures to smooth over gaps.   Periodic behaviors in irregular cadence light curves are sought by interpreting various periodograms:  a generalized Fourier periodogram for sinusoidal signals (Scargle 1982), phase dispersion minimization (Stellingwerf 1978) and a Bayesian procedure for arbitrary shapes (Gregory \& Loredo 1992), and specialized procedures for box-shaped planetary transits (Kovacs et al. 2002, Caceres et al. 2019).

Two critical difficulties in using these methods, and extracting associated scalar features for tabulation and classification, arise when they are applied to sparse, irregular cadenced time series like the LSST WFD survey.  First, the coverage may miss many of the intrinsic characteristics of the variability. Survey observations of a flaring star, for example, might capture a portion of a single flare, missing the peak amplitude and the underlying distribution of smaller and larger flares.  The survey might not capture any flares for a considerable period, so that the star would be misclassified as a quiescent star.  Second, the statistical methods developed for irregular cadences are typically not mathematically well-founded and the statistical measures do not have known asymptotic distributions.  Assumptions like 'independent and identically distributed' observations that underlie the Central Limit Theorem that allow inference of populations from samples often do not apply. It is thus difficult to estimate significant levels for, say, the presence of outliers or periodic variability in an irregularly spaced light curve.   

The Rubin Observatory LSST WFD survey will produce the largest (estimated $\sim 37$ billion) ensemble of sparse and irregular cadence light curves in history.  Bellm (2021) outlines the planned statistical measures that will be automatically produced for each light  curve.  The baseline plan includes simple statistics (flux quantiles, median absolute deviation, skewness and kurtosis, Stetson's $J$ and $K$, normalized excess variance), periodic features (sinusoidal fit periodogram peaks and estimated False Alarm Probabilities), stochastic features (structure function, damped random walk model amplitude), and transient measures. Other measures from the astronomical literature will probably be added.  A few available methods are multivariate, treating the asynchronous multiband observations characteristic of the LSST wide-field survey (Hu \& Tak 2020, Elorrieta et al. 2021, Edes-Huyal et al. 2021). 

\section{Time Series Methods for Dense, Regular Cadences}
\label{anal.sec}

From a statistical viewpoint, the advantage of the LSST single night project is that the single-band evenly-spaced cadence mimics the typical time series widely analyzed in other fields.   The methods have been developed over the past half-century by professional methodologists (statisticians, engineers and economists) to study terrestrial and human-generated time domain phenomena, resting on solid mathematical foundations developed in hundreds of papers in dozens of journals.  The methods are described in texts such as Box et al. (2015), Enders (2015), Hyndman et al. (2018), and Chatfield \& Xing (2019).  Code implementations for hundreds of methods are available in the public domain R statistical software environment (R Core Team 2022); see the Web page {\it CRAN Task Views: Time Series Analysis}.  Many of these methods have strong  foundations in theorems of mathematical statistics and are very widely used. The Box et al. textbook, for example, has over 50 thousand citations, ARIMA modeling has $\sim 2$ million Google hits, and R has $\sim 200$ CRAN packages devoted to regularly cadenced time series.  

Time series analysis of regularly cadenced data often involves multistage procedures such as kernel or Gaussian Processes smoothing, autoregressive modeling, wavelet analysis with thresholded denoising, Fourier analysis with smoothing and tapering, as well as modern signal processing techniques like Hilbert-Huang transform and Singular Spectrum Analysis. Multivariate methods can treat multi-band light curves with methods like VARIMA and dynamic time warping.  These can be combined with a variety of time series diagnostics that are inherently scalar such as probabilities of statistical tests.  A suite of methods for `change point analysis' (Tartakovsky et al. 2014), often coupled with other approaches, can be added to search for sudden changes in temporal behaviors, as occurs in flaring and accretion stellar systems.  These are all coded in R's CRAN packages. 

We address here examples of analysis that is restricted to regular cadenced time series\footnote
{Some gaps in the observations can generally be treated with these methods.  Some procedures permit `missing data' and others can be applied after 'imputation' of missing data.  Sophisticated imputation methods are available that propagate observed variability characteristics into the cadence gap.}.  
The proposed single night micro-survey (\S\ref{1night.sec}) will produce roughly a million stellar LCs, the great majority of which will be consistent with constant flux during the single night observation.  Those which are variable will show a wide range of characteristics (\S\ref{rapidsky.sec}).   While some individual cases might be studied individually $-$ the brightest stars in a given class, stars with anomalous behaviors $-$ the huge quantity of results require that most of the analysis will be based on scalar 'features'  tabulated along with survey catalog information such as astrometry and photometry. All of the standard LSST survey time series features should also be available (Bellm 2021).  These combined features can then be subject to ensemble statistical analysis.

An early step is to determine whether the behavior is stationary or nonstationary.  Nonstationarity includes trends in brightness, periodic or quasi-periodic variations, and short-duration events such as flares and eclipses.   Most stellar variable light curves  will be nonstationary.  The widely used augmented Dickey-Fuller (ADF) test gives a probability of nonstationarity; the  Kwiatkowski-Phillips-Schmidt-Shin test plays a similar role (Kleiber et al. 2008).   A common procedure to remove many forms of nonstationarity is to apply the differencing operator where $x_i$ is replaced with $x_i - x_{i-1}$ to the time series, thereby removing most non-eruptive types of nonstationarity.  

The differenced time series can now be subject to a variety of statistical tests, each revealing some characteristic.  The ADF test can show whether nonstationary behaviors are still present.  The Anderson-Darling test for normality can determine if the noise is Gaussian, and the  Durbin-Watson test can determine if serial (lag=1 cadence interval) autocorrelation is present.   Engle's ARCH test for heteroscedasticity (volatility) and the Brock-Dechert-Scheinkman test for nonlinear or chaotic time series can be applied.

A central issue is to characterize the autocorrelation of both the original and differenced light curve. The question is whether there is stationary autocorrelation, typically some form of short-memory jitter that is superposed on any trend.  This may be astrophysically interesting; both magnetic activity and turbulent accretion disks are expected to have autocorrelated emission.  The autocorrelation function may have dozens of values for different lags and is not immediately amenable to automated inference for characterization or classification.  But  theorem-based tests can give scalar significance levels. The portmanteau Ljung-Box test can discriminate white noise from autocorrelated noise for a range of lags.  The Breusch-Pagan test can give more refined information such as whether significant autocorrelation is present on one minute, ten minutes or one hour timescales.   If a light curve passes these tests, then it has no structure beyond only Gaussian white noise. If it does not pass the tests, the amplitude of variations on different timescales can be quantified with elements of the partial autocorrelation function.

Autoregressive modeling, commonly called ARIMA, can also give illuminating insight into variability.  Here the autoregressive component ('AR') relates current brightness values to recent past values, the integration step ('I') represents the differencing operator to reduce nonstationarity, and the moving average component ('MA') relates current brightness to recent past changes in brightness.  ARIMA models are linear, low dimensional models fit by maximum likelihood estimation with model parsimony established using the Akaike Information Criterion. If the models are successful, as measured with Ljung-Box and ADF tests applied to the ARIMA residuals, then detailed information into the variability characteristics are embedded in a few coefficients.  

There are many extensions to ARIMA: ARFIMA adds a power law $1/f^\alpha$ long-memory component that astronomers call 'red noise'; VARIMA treats multivariate regular cadence time series with lags; GARCH (development of which was awarded the 2003 Nobel Prize in Economics) added stochastically varying variances (volatility).   

ARIMA modeling is commonly used in signal processing and econometrics but only rarely in astronomy because of the sparse and irregular cadences (Feigelson et al. 2018).  Application to variable stars in NASA's Kepler 4-year mission shows that most (but not all) cases are well fit with low-dimensional ARIMA models (Caceres et al. 2019b).  

Fourier analysis is clearly important to identify stellar variability with strict periodic components from orbital, rotational or pulsational properties.  Here a long-established methodology is available for constructing, improving, and interpreting power spectra; see texts like Priestley (1982) and Percival \& Walden (1993).  Extraction of reliable scalar features is not trivial, as power spectra often benefit from carefully crafted improvements such as multitapering and smoothing.  Alias structures are commonly present when true periodicities are present.  Nevertheless, with sufficiently careful techniques, clear periodic signals can often be identified in an automated fashion. 

\section{Ensemble Analysis of Regular Cadence Light Curves}
\label{ensemble.sec}

The classification of variable stars is rooted in historical studies of the mid-20th century.  Scholars like Cecelia Payne and Boris~V. Kukarkin invented dozens of classes like SS~Cyg, R~Cor~Bor, W~U~Ma and BY~Dra stars based on erratic pre-CCD photometric measurements and subjective judgment of variability behaviors (Payne-Gaposchkin et al. 1938, Kukarkin et al. 1969).  There is no objective taxonomy of variable stars based on quantitative statistical analysis of identically cadenced LCs obtained under similar conditions.  This Rubin Observatory micro-survey has the potential of rectifying this gap for rapid stellar variability, clarifying which historical classes are distinct, quantifying their occurrence rates, and discovering new classes that have escaped identification. 

Automated processing could produce around 40 useful scalar features for ensemble studies.  These might include around 10 measures from the standard LSST time series processing (Bellm 2021), 20 features from the analysis outlined above (\S\ref{anal.sec}), and 10 metadata features (including single-epoch 6-band photometry) from the standard LSST source catalog.  Many of these features have been used in past analyzes of large ensembles of light curves such as the Kepler satellite and Zwicky Transient Facility survey (Caceres et al. 2019b, Coughlin et al. 2021). If 10-20\% of the stars exhibit some form of variability, the LSST single night micro-survey outlined in \S\ref{1night.sec} would then produce a multivariate database with $\sim 100,000-200,000$ rows and $\sim 40$ columns.  

This database would contain classes of variable stars outlined in \S\ref{rapidsky.sec}, slow trends from variables with characteristic timescales $>1$ day, and outliers that may have either instrumental or astronomical origins.  Identification of these classes, and identifying likely members of each class, requires statistical $clustering$ procedures.  Statistical $classification$ procedures can only be applied if expert astronomers manually label a sufficient number of variable stars to provide training sets.  Full training sets are probably too difficult to obtain for the faint magnitudes that will dominate the LC ensemble, but semi-supervised clustering guided by partial training sets can be pursued (Basu et al. 2008).  

Classical nonparametric clustering methods $-$ such as hierarchical clustering, $k$-means partitioning, and Voronoi tessellations $-$  are unlikely to provide reliable classes in this high-dimensional space.  Even for simple problems, the classes are often unstable and depend on arbitrary choices of method and parameters (Everitt et al. 2011).  For example, single-linkage clustering (known in astronomy as the `friends-of-friends' algorithm) is both subject to spurious `chaining' and is computational inefficient with $O(N^3)$.  Parametric methods, such as Gaussian mixture models, can give more stable results and probabilities of membership for each star in each cluster (Bouveyron et al. 2019).   Efficient implementations of hierarchical clustering are given in CRAN $fastcluster$ package in the R software environment, but results may not be effective for high dimensions. 

While these traditional clustering procedures can be tried, more innovative and computationally efficient methods may be needed. Several hybrid algorithms are widely used for clustering large datasets including CLARA, CURE, BIRCH, and HDBSCAN (Wunsch \& Xu 2008). Other exploratory approaches include the CRAN package $scagnostics$ developed by distinguished computer visualization scholar Leland Wilkinson.  Here unusual scatterplots are identified from a high-dimensional dataset using graph theoretic measure (Wilkinson 2005).   Wilkinson's $HDoutliers$ package produces clusters and identifies outliers by distributed aggregation; it can reduce a $10^8$-object dataset in less than a minute (Wilkinson 2018).  Outliers can also be sought from visualization of multivariate projections onto two dimensions using the t-SNE algorithm (Giles et al. 2019). Another interesting possibility is the Statistical Information Grid-based (STING) clustering procedure that divides the $p$-space into a multi-resolution hierarchy of rectangular cells, providing only simple counts in the cells (Wang et al 1997). STING has sub-linear computational complexity. 

The single night LSST micro-survey also has the potential to discover unexpected rapid photometric events, both stellar and extragalactic.  A variety of methods have been developed for finding anomalies in multivariate datasets without a clear winner (Chandola 2009).  Some procedures start with a $O(N)$ fixed-width clustering technique or $kd$ trees with additional anomaly detection rules. The Isolation Forest algorithm has gained popularity due to high computational efficiency and strong ability to find clumps of objects with unusual properties (Liu et al. 2008).  Several modern distance- and density-based outlier algorithms are provided in CRAN package {\it DDoutlier}, and earlier outlier detection methods are available in CRAN packages {\it FastPCS}, {\it mvoutlier} and {\it Routliers}.  

~\\

\section{Design of a Single Night LSST Micro-Survey }
\label{1night.sec}

An LSST field at low Galactic latitude will be observed continuously for a long winter night ($\sim 12.5$ hr) with $\sim 1500$ regularly spaced 30 sec exposure in the $r$ band.   An additional single-epoch 60 min observation of the same field on a different night is required to obtain 10~min photometry in all six Rubin Observatory photometric bands.  This will allow magnitudes and colors to serve as metadata to supplement variability characteristics for statistical classification of the objects.   This micro-survey would be best  conducted during Rubin Observatory commissioning phase to avoid interrupting planned cadence for the main WFD survey and to provide guidance for classification of WFD variable stars. 

The exact field location would be chosen to give a high stellar density without significant crowding, little saturation by bright stars,  low interstellar absorption, a mixture of Population I and II stars, and long horizon-to-horizon winter coverage at the Rubin Observatory site.  Consider a target field around $(l,b) = (330^\circ, +20^\circ)$.  A stellar population synthesis simulation from the Besan\c{c}on model (Robin et al. 2003) predicts that this 9.6 deg$^2$ LSST field  with a sensitivity limit $r < 24.0$ magnitude will have $\sim 1.2$ million stars. About 700,000 have $r<22$ to give high S/N LCs. Most stars are at distances between 1 and 8 kpc with ages $3-10$ Gyr;  median absorption $A_V \sim 0.2$ mag. Most of the targets are K and M stars with $\sim 30,000$ F-G stars. 

After standard reduction of the images by the LSST Data Management pipeline, and production of standard LSST tabular results with astrometry and photometry, time series analysis on the $\sim 1500$ regularly spaced measurements would be performed.  An estimate of the computation burden would be 0.1 CPU-hour for each star or $\sim 100,000$ CPU-hours for the full sample.  Science analysis (e.g., clustering and classification) would then proceed with an interactive process of statistical clustering procedures and visual inspection of selected light curves.  The computational effort would be light during the exploratory phase using moderate-size subsamples ($10^4$ stars) and heavier when the final analysis is performed on the full sample.  Much of the analysis is `embarrassingly parallel' so an arbitrary number of cores can operate independently. 

Statistical and scientific analysis would be performed in the R statistical software environment where most of the methods are implemented (R Core Team 2022).  Typically, computationally intensive stages of these R programs are typically coded in C++.  The final script can be easily wrapped into Python for use by the wider community.  

Finally, the analysis of these rapid and regular cadence light curves should inform the analysis of variable objects in the sparse and irregular cadenced WFD and Deep Drilling Field observations.   The clustering study will have a dimensional reduction stage where some features are found to be extraneous and others are effective in discriminating classes.  The most useful features can then guide statistical characterization of the irregular LCs, speeding classification as the all-sky survey progresses. These results may be of particular interest to LSST Event Brokers that will try to classify variable stars as the surveys progress.

\section{Summary}
\label{summary.sec}

The principal Rubin Observatory LSST Wide Fast Deep and Deep Drilling Field multiband surveys will detect billions of variable cosmic objects.  But their cadences will be irregular and sparse so that rapid variability -- on timescales of minutes to hours -- will be poorly characterized.  Here we propose a rapid cadence micro-survey during a single night, observing a single field at a single band designed to capture rapidly varying stellar phenomena.  About a million light curves will be obtained.  These include flaring late-type stars, short-period binary systems and cataclysmic variables, possibly young stellar objects and ultra-short period exoplanets, and unknown anomalous behaviors. The number of rapidly variable objects is largely unknown; this single night survey will give unique insights into this regime of transient astronomy.

A large suite of well-established statistical and machine learning procedures can be brought to analyze this unique collection of regularly spaced light curves. These methods, developed over decades for signal processing and econometrics, are not applicable to sparse and irregular cadences.  They will give dozens of scalar features to characterize in detail different modes of variability.  These features can then be subject to advanced unsupervised (or semi-supervised) clustering  procedures to give a unique, authoritative, and objective classification of rapidly variable stars.  

The result can be the first $objective$ taxonomy of rapidly variable stars based on quantitative statistical analysis of identically cadenced LCs obtained under similar conditions.  In essence, this one-night observation can provide the gold standard of variability studies on short timescales.

The most effective features can then inform the wider Rubin Observatory community on the best approaches to variable star identification and classification from the sparse, irregular cadences that dominate the planned surveys.  The findings will be available for the full LSST community, including Event Brokers, to improve classification for the dominant irregularly cadenced surveys.

~\\

Acknowledgements: We appreciate the thoughtful comments of an anonymous referee.

\vfill\newpage

Andreoni et al., 2020, MNRAS, 491, 5852

Astropy Collaboration et al., 2018, AJ 156, 123,

Basu, S., Davidson, I., \& Wagstaff, K.~e. 2008, Constrained Clustering:
  Advances in Algorithms, \\ \hspace*{0.3in} Theory, and Applications (Chapman \& Hall)

Bellm, E. 2021, Review of Timeseries Features, Tech. Rep. DMTN-118, LSST Corp.

Belloni, D., et al., 2018, MNRAS, 478, 5626

Bienias, J., et al. 2021, ApJSuppl 256, 11

B{\'o}di, A., \& Hajdu, T. 2021, ApJSuppl 255, 1

Bonito, R., et al., 2023, ApJSuppl 265, 27

Bouveyron, C., Celeux, G., Murphy, T.~B., \& Raftery, A.~E. 2019, Model-Based
  Clustering and  \\ \hspace*{0.3in} Classification for Data Science (Cambridge University Press)

Box, G. E.~P., Jenkins, G.~M., Reinsel, G.~C., \& Ljung, G.~M. 2015, Time
  Series Analysis:   \\ \hspace*{0.3in} Forecasting and Control, 5th edn. (Wiley)

Breivik, K., et al. 2022.
https://www.overleaf.com/project/6244a25dfefc2c4cf76d841f

Brown, W.~R.,  {et~al.} 2020, ApJ 889, 49

Burdge, K.~B., et al. 2020, ApJ 905, 32,

Caceres, G.~A., {et~al.}, 2019, AJ 158, 57

Caceres, G.~A., {et~al.}, 2019b, AJ 158, 58

Chand, K., {et~al.} 2022, MNRAS 511, 13

Chandola, V., Banerjee, A., \& Kumar, V. 2009, ACM Comp. Surveys, 31.  \\ \hspace*{0.3in}
https://dl.acm.org/doi/10.1145/1541880.1541882

Chatfield, C., \& Xing, H. 2019, The Analysis of Time: An Introduction with R,
  7th edn.  \\ \hspace*{0.3in} (CRC Press)

Chen, X.,  {et~al.} 2020, ApJSuppl 249, 18

Coughlin, M.~W., {et~al.} 2021, MNRAS 505, 2954

Davenport, J. R.~A. 2016, ApJ 829, 23

Dobrotka, A., et~al. 2021, As\&Ap 649, A67

Drake, A.~J., {{et~al.} 2014, MNRAS 441, 1186

Edelson, R.~A., \& Krolik, J.~H. 1988, ApJ, 333, 646

Edes-Huyal, F.~K., et~al. 2021, AJ 161, 168

Elorrieta, F., et al. 2021, MNRAS 505, 1105

Enders, W. 2014, Applied Econometric Time Series, 4th edn. (Wiley)

Everitt, B.~S., Landau, S., Leese, M., \& Stahl, D. 2011, Cluster Analysis, 5th
  edn. (Wiley)

Feigelson, E.~D., et~al. 2018, Frontiers in Physics, 6, 80

Feigelson, E.~D., {et~al.} 2011, ApJSuppl 194, 9

Giles, D., \& Walkowicz, L. 2019, MNRAS 484, 834

Graham, M.~L., {et~al.} 2023, MNRAS 519, 3881

Gregory, P.~C., \& Loredo, T.~J. 1992, ApJ 398, 146

Gullbring, E. 1994, As\&Ap, 287, 131

G{\"u}nther, M.~N.,  {et~al.} 2020, AJ 159, 60

Harris, A.~W., \& {Pravec}, P. 2006, in Asteroids, Comets, Meteors, ed.
  D.~{Lazzaro} et al., Vol. 229,  \\ \hspace*{0.3in} 439--447

{Hawley}, S.~L., {et~al.} 2014, ApJ 797, 121

{Hawley}, S.~L., \& {Pettersen}, B.~R. 1991, ApJ 378, 725

{Hu}, Z., \& {Tak}, H. 2020, AJ 160, 265

Hyndman, R., \& Athanasopoulos, G. 2018, Forecasting: Principles and Control,
  2nd edn.  \\ \hspace*{0.3in} (OTexts)  https://otexts.com/fpp2/

{Ivezi{\'c}}, {\v{Z}}., et al. 2019, ApJ 873, 111

Kleiber, C., \& Zeileis, A. 2008, {Applied Econometrics with R} (Springer)

{Kov{\'a}cs} et al. 2002, As\&Ap 391, 369

{Kowalski}, A.~F., {et~al.} 2009, AJ 138, 633

{Kruckow}, M.~U., et al. 2021, ApJ 920, 86

{Kukarkin}, B.~V., {et~al.} 1969, {General
  Catalogue of Variable Stars. Volume\_1. Constellations  \\ \hspace*{0.3in} Andromeda - Grus.}

{Kulkarni}, S. 2006, in KITP Conference: Transient Universe: Popular, Not so
  Popular \& Knowable  \\ \hspace*{0.3in} Unknowns, ed. L.~{Bildsten} et al., 14

{Kulkarni}, S.~R. 2012, arXiv:1202.2381

Liu, F.~T., et al. 2008, IEEE Conf. Data Mining '08

{LSST Science Collaboration}, {Abell}, P.~A., {et~al.} 2009,
arXiv:0912.0201.

{Maehara}, H., {et~al.} 2015, Earth, Planets and
  Space, 67, 59

{Maehara}, H., {et~al.} 2012, Nature, 485, 478,

{McCormac}, J., {et~al.} 2020, MNRAS 493, 126

{Mowlavi}, N., {et~al.} 2018, As\&Ap 618, A58

{Mullan}, D.~J., \& {Houdebine}, E.~R. 2020, ApJ 891, 128,

{Namekata}, K., {et~al.} 2017, ApJ 851, 91

{Ofek}, E.~O., {et~al.} 2020, MNRAS 499, 5782

{Ojha}, V., et al. 2022, MNRAS

{Pala}, A.~F., {et~al.} 2020, MNRAS 494, 3799

{Payne-Gaposchkin}, C., \& {Gaposchkin}, S. 1938, {Variable stars}

Percival, D.~B., \& Walden, A.~T. 1993, Spectral Analysis for Physical
  Applications: Multitaper  \\ \hspace*{0.3in} and Conventional Univariate Techniques (Cambridge
  University Press)

Priestley, M.~B. 1982, Spectral Analysis and Time Series, vols. i and ii edn.
  (Academic Press)

{R Core Team}. 2021, R: A Language and Environment for  \\ \hspace*{0.3in} Statistical Computing, R
  Foundation for Statistical Computing, Vienna, Austria.  \\ \hspace*{0.3in} 
https://www.R-project.org/  

{Robin}, A.~C., et al., S. 2003, As\&Ap 409, 523

{Saha}, A., {et~al.} 2019, ApJ 874, 30

{Scargle}, J.~D. 1982, ApJ 263, 835

{Schaefer}, B.~E., {et~al.} 1987, ApJ 320, 398

{Schmidt}, S.~J., {et~al.} 2019, ApJ 876, 115

{Shields}, A.~L., et al. 2016, Physics Reports, 663, 1,

{Smith}, A.~M.~S., {et~al.} 2018, MNRAS 474, 5523

{Smith}, K.~W., {Jones}, D.~H.~P., \& {Clarke}, C.~J. 1996, MNRAS, 282, 167,

{Soszy{\'n}ski}, I., {et~al.} 2008, Acta Astronomica, 58, 293.

{Stellingwerf}, R.~F. 1978, ApJ 224, 953

{S{\"u}veges}, M., {et~al.} 2015, MNRAS 450, 2052

{Tu}, Z.-L., et al. 2021, ApJSuppl 253, 35

{Tyson}, J.~A. 2002, in SPIE Conf. Ser. \#4836, Survey and Other Telescope Technologies
  and  \\ \hspace*{0.3in} Discoveries,  ed. J.~A. {Tyson} \& S.~{Wolff}, 10--20

{VanderPlas}, J.~T. 2018, ApJSuppl 236, 16

{Venuti}, L., {et~al.} 2021, AJ 162, 101

Wang, W., J., Y. \& Muntz, R. 1997, 23rd Intl. Conf. on Very Large Data Bases,
  18.   \\ \hspace*{0.3in}  https://dl.acm.org/doi/10.5555/645923.758369

{Webb}, S., {et~al.} 2021, MNRAS 506, 2089

{Welch}, D.~L., \& {Stetson}, P.~B. 1993, AJ 105, 1813

Wilkinson, L. 2018, IEEE Trans. Vis. Comp. Graphics, 24.

Wilkinson, L., Anand, A., \& Grossman, R. 2005, IEEE Symp. Information Vis., 21.

{Worden}, S.~P., et al. 1981, ApJ 244, 520

Wunsch, D.~C., \& Xu, R. 2008, Clustering (Wiley - IEEE Press)

\end{document}